\documentclass[aps,prl,twocolumn,floatfix]{revtex4}
\usepackage{eurosym}
\usepackage{amsfonts}
\usepackage{mathrsfs}
\usepackage{color}
\usepackage{xspace}
\usepackage{subfigure}
\usepackage{longtable}
\usepackage{xspace}
\usepackage{multirow}
\usepackage{tabularx}
\usepackage{amsmath}
\usepackage{amssymb}
\usepackage{graphicx}
\usepackage{dcolumn}
\usepackage{bm}
\usepackage[colorlinks,urlcolor=blue,citecolor=blue,linkcolor=blue]{hyperref}
\usepackage[normalem]{ulem}
\usepackage{epstopdf}

\begin{document}

\title{Spin-momentum coupled Bose-Einstein condensates with lattice band
pseudospins}
\author{M. A. Khamehchi$^{1}$}
\author{Chunlei Qu$^{2}$}
\author{M. E. Mossman$^{1}$}
\author{Chuanwei Zhang$^{2}$}
\thanks{chuanwei.zhang@utdallas.edu}
\author{P.~Engels$^{1}$}
\thanks{engels@wsu.edu}

\begin{abstract}
The quantum emulation of spin-momentum coupling (SMC), a crucial ingredient
for the emergence of topological phases, is currently drawing considerable
interest. In previous quantum gas experiments, typically two atomic
hyperfine states were chosen as pseudospins. Here, we report the observation
of a new kind of SMC achieved by loading a Bose-Einstein condensate (BEC)
into periodically driven optical lattices. The $s$- and $p$-bands of a
static lattice, which act as pseudospins, are coupled through an additional
moving lattice which induces a momentum dependent coupling between the two
pseudospins, resulting in $s$-$p$ hybrid Floquet-Bloch bands. We investigate
the band structures by measuring the quasimomentum of the BEC for different
velocities and strengths of the moving lattice and compare our measurements
to theoretical predictions. The realization of SMC with lattice bands as
pseudospins paves the way for engineering novel quantum matter using
hybrid orbital bands.
\end{abstract}

\affiliation{$^1$Department of Physics and Astronomy, Washington State University,
Pullman, WA 99164, USA \\
$^{2}$Department of Physics, The University of Texas at Dallas, Richardson,
Texas 75080, USA}

\maketitle

Spin-momentum coupling (SMC), commonly called spin-orbit coupling, is a
crucial ingredient for many important condensed matter phenomena such as
topological insulator physics, topological superconductivity, spin Hall
effects, etc~\cite{Zutic2004,Hasan2010,Qi2011}. In this context, the recent
experimental realization of SMC in ultracold atomic gases provides a
powerful platform for engineering many interesting and novel quantum phases~%
\cite{Lin2011,Pan2012,Wang2012,Cheuk2012,Hamner2014,Olson2014}. In typical
experiments, two atomic hyperfine states act as two pseudospins which are
coupled to the momentum of the atoms through stimulated Raman transitions~%
\cite{Higbie2002,Spielman2009}. However, ultracold atoms in optical lattice
potentials possess other types of degrees of freedom which can also be used
to define pseudospins~\cite{Jaksch1998,Bloch2005}. A natural and important
question is whether such new types of pseudospins can be employed to
generate SMC.

In optical lattices filled with ultracold atoms, \textit{s}- and \textit{p}%
-orbital bands are separated by a large energy gap and can be defined as two
pseudospin states. One significant difference between hyperfine state
pseudospins and lattice band pseudospins lies in the energy dispersion of
\textquotedblleft spin-up\textquotedblright\ and \textquotedblleft
spin-down\textquotedblright\ orientations: the dispersion relations are the
same for hyperfine state pseudospins, while they are inverted for lattice
band pseudospins. It is well known from topological insulators and
superconductor physics that inverted band dispersions, together with SMC,
play a central role for topological properties of materials~\cite%
{Bernevig2006,Konig2007,Fu2008}. Therefore, it is natural to expect that the
inverted band pseudospins, when coupled with the lattice momentum, may lead
to interesting topological phenomena in cold atomic optical lattices. Recent
experiments with shaken optical lattices (i.e. lattices in which the lattice
sites are periodically shifted back and forth in time~\cite{Parker2013})
have realized a simple coupling ($\Omega \sigma _{x}$ coupling, where $%
\Omega $ is the coupling strength and $\sigma _{x}$ a Pauli matrix) between
\textit{s}- and \textit{p}-band pseudospins, analogous to Rabi coupling
between two regular spins~\cite{Zheng2014}. However, for the exploration of
exotic phenomena in optical lattice systems, such as
Fulde-Ferrell-Larkin-Ovchinnikov (FFLO) phases~\cite{FF64,LO64} and Majorana
fermions~\cite{Fu2008}, SMC with $s$- and $p$-bands pseudospins is highly
desirable~\cite{Zheng2013,Wu2013,Zhang2008,Sato2009}.

In our experiments we realize such $s$-$p$ band SMC for a Bose-Einstein
condensate (BEC) using a weak moving lattice to generate Raman coupling
between \textit{s}- and \textit{p}-band pseudospins of a static lattice~\cite%
{Muller2007}. The moving lattice acts as a periodic driving field \cite%
{Lignier2007,Eckardt2010,Hauke2012,Goldman2014a,Goldman2014b,Dalessio2014} and has
previously been used to generate an effective magnetic field in the lowest $s
$-band of a tilted optical lattice \cite{Aidelsburger2013,Miyake2013}. In
our experiment, the driving frequency of the moving lattice is chosen close
to the energy gap between $s$- and $p$-bands at zero quasimomentum, leading
to a series of hybrid $s$-$p$ Floquet-Bloch (FB) band structures. FB band
structures in optical lattices give rise to interesting and important
phenomena in cold atoms and solids~\cite{Sengstock2011,Struck2013}, as is
evidenced by the recent experimental realization of a topological Haldane
model in a shaken honeycomb optical lattice \cite{Jotzu2014} and the
observation of FB states on the surface of a topological insulator~\cite%
{Wang2013}.

Here we show that the moving lattice generates two types of coupling between
\textit{s-} and \textit{p}-band pseudospins: a momentum-independent Rabi
coupling ($\Omega \sigma _{x}$) and SMC ($\alpha \sigma _{x}\sin (q_{x} d) $%
, where $q_{x}$ is the quasimomentum and $d$ the lattice period), with
strengths of the same order. The coexistence of these two types of coupling
leads to asymmetric FB band dispersions~\cite{Zheng2015}. We investigate the
FB band structures by measuring the quasimomentum of the BEC. The initial
phase of the moving lattice plays a significant role in the Floquet dynamics
\cite{Goldman2014a}, the effects of which are explored through a quantum
quench induced dynamical coupling of the FB bands. Results are compared to
theoretical predictions from a simple two-band model and from numerical
simulations of the Gross-Pitaevskii (GP) equation.\newline

\noindent{\Large {\textbf{Results}}}

\begin{figure}[t!]
\centering\includegraphics[width=0.48\textwidth]{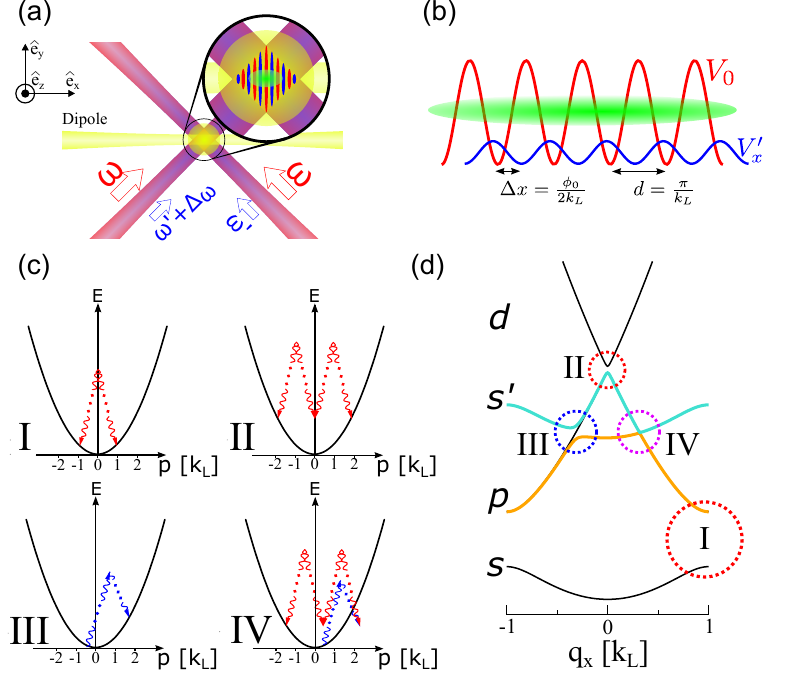}
\caption{\textbf{Experimental setup and schematic lattice illustration.} (a)
Experimental arrangement. The crossed dipole trap beams propagate in the $%
\vec{e}_x$ and $\vec{e}_z$ direction. The static and moving lattice have
overlapping beams propagating along $\vec{e}_x + \vec{e}_y$ and $-\vec{e}_x
+ \vec{e}_y$. (b) Lattice potentials along the $\vec{e}_x$ direction. The
lattice period $d$ is identical for the static lattice $V_0$ and the moving
lattice $V^{\prime }_x$. The initial offset between lattice sites of the
static and moving lattice, $\Delta x$, is given by the initial phase $%
\protect\phi_0$ between the two lattices. (c,d) Illustration of the
multi-photon processes for the driven lattice system and the corresponding
FB band structure in the first Brillouin zone. The static lattice induces a
large energy gap (I) through a 2-photon process and a small energy gap (II)
through a 4-photon process. The moving lattice induces an energy gap when
the $s$-band and $p$-band are coupled through (III). A smaller energy gap is
produced by a combination of the static and moving lattice (IV).}
\label{schematic}
\end{figure}

\noindent\textbf{Experimental setup}. To generate the $s$-$p$ band SMC and
FB band structures, we begin with a $^{87}$Rb BEC composed of approximately $%
5\times10^{4}$ atoms confined in a crossed dipole trap. A static lattice is
generated by two perpendicular laser beams with wavelength $\lambda \approx
810$~nm intersecting at the position of the BEC, as schematically shown in
Fig.~\ref{schematic}(a). The harmonic trap frequencies due to the envelope
of the static lattice beams and the crossed dipole trap are $(\omega
_{x},\omega _{y},\omega _{z})=2\pi \times (41,159,115)$~Hz, where $\vec{e}%
_{x}$ points along the lattice, $\vec{e}_{y}$ is the horizontal transverse
direction, and $\vec{e}_{z}$ is the vertical direction. A weak moving
lattice with the same lattice period as the static lattice, $d=\pi /k_{L}$
where $k_{L}=\sqrt{2}\pi /\lambda $, is then overlaid with the static
lattice (Fig.~\ref{schematic}(b)). The moving lattice beams are
approximately $180$~MHz detuned from the static lattice. A small frequency
difference $\Delta _{\omega }$ between the two moving lattice beams
determines the velocity of the lattice according to $v_{lattice}=\Delta
_{\omega }/2k_{L}$. To induce $s $-$p$ orbital band coupling, $|\Delta
_{\omega }|$ is chosen close to the energy gap $E_{sp}$ between the $s$- and
$p$-bands of the static lattice at quasimomentum $q_{x}=0$.

One outstanding feature of the coupling scheme employed in these experiments
is the asymmetry of the effective $s$-$p$ FB bands, which exhibit a local
minimum located at a finite quasimomentum $q_x \neq 0$. The direction in
which the minimum is shifted away from $q_x=0$ is determined by the sign of $%
\Delta_\omega$ (which determines the direction of motion of the moving
lattice) and $|\Delta_\omega|-E_{sp}$ (i.e. the detuning of the drive from
the bandgap at $q_{x}=0$). Before describing experimental results and a
formal derivation of the band structure using Floquet theory~\cite%
{Goldman2014a, Goldman2014b}, we lay the groundwork by presenting a
multi-photon resonance picture that provides intuitive insights (Fig.~\ref%
{schematic}(c,d)). In this picture, one starts with the parabolic dispersion
of a free atom in the absence of any external potentials. An optical lattice
then induces $2n$-photon couplings (with $n$ being an integer number)
between points of the dispersion relation due to absorption and stimulated
emission processes. The couplings are centered around pairs of points that
fulfill conservation of energy and momentum. At these points, bandgaps open
due to avoided crossings. Examples for possible couplings due to the static
lattice (red arrows in Fig.~\ref{schematic}(c)) and the moving lattice (blue
arrows in Fig.~\ref{schematic}(c)) and the associated bandgaps in the first
Brillouin zone are shown in Fig.~\ref{schematic}. Different coupling
strengths lead to different sizes of bandgaps, which result in an asymmetric
band structure.

In another pictorial way, the Floquet band structure for the time-periodic
system can be constructed by creating multiple copies of the Bloch band
structure of the static lattice that are offset in energy by $|\Delta_\omega|
$. The moving lattice couples the $p$-band and the shifted $s$-band
(labelled by $s^\prime$ in Fig.~\ref{schematic}(d)) at points where the
shifted $s$-band intersects the unshifted $p$-band. The gaps opened by the
coupling can formally be calculated using Floquet theory.

\noindent\textbf{Experimental measurements}. Adiabatic loading of the BEC
into an $s $-$p$ FB band is achieved by first ramping on the intensity of
the static lattice, followed by adiabatically ramping on the moving lattice
intensity. In this procedure, the initial relative phase between the two
lattices, $\phi_0$ (Fig.~\ref{schematic}(b)), becomes irrelevant and can
effectively be set to zero. As we shall show in the context of Fig.~\ref%
{phase}, if the moving lattice is suddenly jumped on instead of
adiabatically ramped on, this initial relative phase may manifest itself by
drastically changing the dynamics of the system~\cite{Goldman2014a}.

\begin{figure}[t!]
\centering\includegraphics[width=0.48\textwidth]{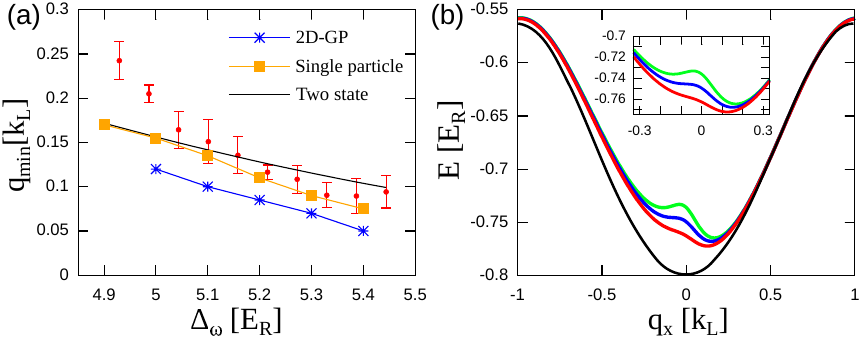}
\caption{\textbf{Effects of the driving frequency.} (a) Band minimum $q_{min}
$ for the upper hybrid band vs. driving frequency $\Delta _{\protect\omega }$%
. The depth of the moving lattice is $1~E_{R}$. The filled circles are
experimental measurements. The black line shows the theoretical prediction
of a two-band model. The squares and stars are the results of numerical
simulations of the Schr\"odinger equation and the GP equation, respectively.
(b) Upper hybrid $s$-$p$ FB band structure for different driving frequencies
$\Delta _{\protect\omega }=4.99~E_{R}$, $5.1~E_{R}$ and $5.22~E_{R}$ from
top to bottom. The lowest (black) curve is the $s$ orbital band without the
presence of the driving field.}
\label{velocity}
\end{figure}

Figure \ref{velocity}(a) shows the measured position, $q_{min}$, of the band
minimum for different driving frequencies, $\Delta_\omega$, after
adiabatically loading a BEC into a FB band. The driving frequencies are
chosen such that $\hbar \Delta_\omega$ lies in the gap at $q_x = 0$ between
the $p$-band ($4.64~E_R$, where $E_{R}=\hbar^2 k_L^2/{2m}=h\times 1749.5$
Hz) and the $d$-band ($5.44~E_R$). After adiabatically loading a BEC into a
FB band, the lasers are switched off and the BEC is imaged after 14~ms
time-of-flight (TOF). The positional shift of the BEC components is then
used to determine the quasimomentum. Each data point is an average over five
iterations of the measurement. A shift of the quasimomentum is detected that
decreases with increasing driving frequency (Fig.~\ref{velocity}(a)) as the
coupling between the $p$-band and shifted $s$-band becomes weaker. The
observed shift indicates a shift of the minimum of the upper hybrid band
(Fig.~\ref{velocity}(b)) into which the BEC is adiabatically loaded. The
solid line in Fig.~\ref{velocity}(a) shows $q_{min}$ calculated from a
simple two-band model (see below) and is in reasonable agreement with the
data. The symbols are the results from real time simulation of the
Schr\"odinger equation (squares) and the GP equation (stars) with finite
nonlinear interaction strength~\cite{note0}. We see that the interaction
could modify the single-particle results.

\begin{figure}[t!]
\centering\includegraphics[width=0.48\textwidth]{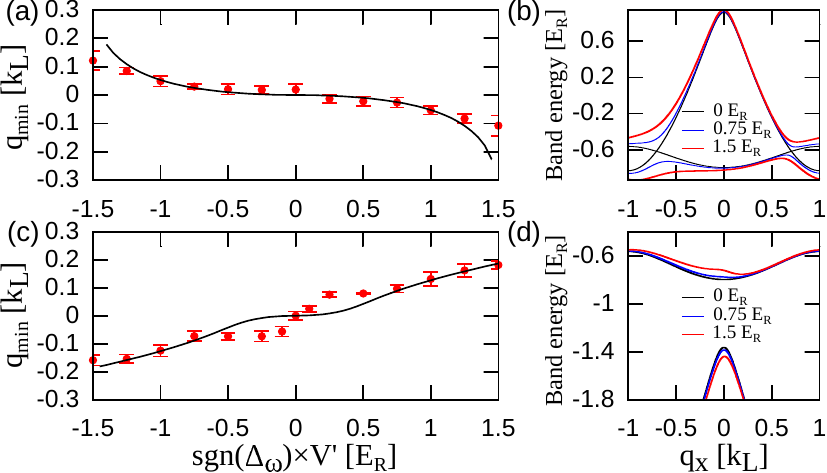}
\caption{\textbf{Effects of the driving strength.} Band minimum $q_{min}$
vs. driving field strength $V^\prime$ for different driving frequencies of
(a) $\lvert\Delta_\protect\omega\rvert=2.92E_R$ and (c) $\lvert\Delta_%
\protect\omega\rvert = 5.21E_R$. The red points are experimental data, the
solid lines are the theoretical predictions from a two-band model. $%
sgn(\Delta\protect\omega)$ determines the direction of motion of the moving
lattice. (b,d) Corresponding hybrid band structures for different driving
field strengths $V_x^\prime = 1.5 E_R$, $0.75 E_R$ and $0 E_R$ (outer to
inner curves).}
\label{strength}
\end{figure}

Figure \ref{strength} presents a complementary data set for which the
driving frequency is set to a constant value with $\lvert\Delta_\omega%
\rvert<E_{sp}$ (Fig.~\ref{strength}a) or $\lvert\Delta_\omega\rvert>E_{sp}$
(Fig.~\ref{strength}c) and the quasimomentum is determined for various
depths of the moving lattice. The sign of $\Delta_\omega$ determines the
direction of motion of the moving lattice. For $\lvert\Delta_\omega%
\rvert<E_{sp}$ the BEC resides in the lower hybrid $s$-$p$ FB band (Fig.~\ref%
{strength}(b)) while for $\lvert\Delta_\omega\rvert>E_{sp}$ it is in the
upper hybrid band (Fig.~\ref{strength}(d)). This leads to a shift of the
quasimomentum into opposite directions for the two cases. For a given
driving frequency, the coupling of the two bands is stronger for larger
driving field strength (i.e. larger depth of the moving lattice) so that the
BEC is shifted to a larger absolute value of quasimomentum.

\begin{figure}[t!]
\centering\includegraphics[width=0.48\textwidth]{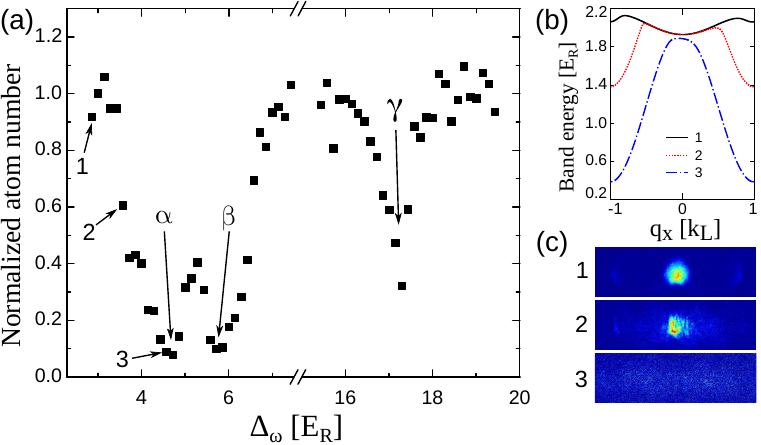}
\caption{\textbf{Heating of the Floquet system.} (a) Number of atoms
remaining after adiabatically loading a BEC into the FB band, normalized to
initial atom number determined from independent experimental runs. The
static lattice is ramped on to $5.47~E_R$ in $200~ms$. Then the moving
lattice is ramped on to a depth of $V^\prime = 0.5~E_R$ in $60~ms$. The dips
$\protect\alpha$, $\protect\beta$, and $\protect\gamma$ occur close to the
Bloch bands $p$, $d$, and $f$. (b) Effective band structures for the lower
hybrid band for data points $1$, $2$, and $3$ of panel (a). (c) TOF images
taken at points $1$, $2$, and $3$.}
\label{stability}
\end{figure}

Floquet systems such as the one in our experiment are described by
quasienergy bands. They do not have a thermodynamic ground state, and in the
presence of many-body interactions their stability can be affected by a
variety of factors~\cite{Choudhury2014a,Choudhury2014b,Cooper2015}.
Experimentally, we study the stability of the system by determining the
number of condensed atoms left after the static and the moving lattices are
successively and adiabatically ramped on. TOF imaging reveals atom loss and
heating of the BEC as shown in Fig.~\ref{stability}. The dips $\alpha $, $%
\beta $, and $\gamma $ in Fig. \ref{stability}(a) occur when the driving
frequency is chosen such that it leads to a coupling close to the Bloch
bands $p$, $d$ and $f$ of the static lattice at $q_{x}=0$ respectively. The
lower hybrid band structure, in which the BEC mainly resides, for points $1$%
, $2$, and $3$ and the corresponding TOF images are shown in panels (b) and
(c). Resonance induced collective excitations and modulational instabilities
can play a role for the observed losses~\cite{JimenezGarcia2014}.

\noindent\textbf{Minimal two-band model}. The dynamics of the BEC are
governed by the full time-dependent GP equation, $i\hbar \frac{\partial }{%
\partial t}\psi (\mathbf{r},t)=[H_{0}(t)+V_{trap}+V_{int}]\psi (\mathbf{r},t)
$ where $V_{trap}$ and $V_{int}$ are the external trapping potential and the
mean-field interaction, respectively. $H_{0}(t)$ is the single-particle
Hamiltonian,
\begin{equation}
H_{0}(t)=\frac{p^{2}}{2m}+V_{0}\cos
^{2}(k_{L}x)+V^{\prime}\cos^{2}(k_{L}x+\phi _{0}-\frac{\Delta _{\omega }t}{2}%
),
\end{equation}
where the second and the third terms describe the static and moving optical
lattices, respectively, and $\phi _{0}$ is the initial relative phase
between the two sets of lattices.

When the static lattice depth $V_{0}$ is large and when $|\Delta _{\omega }|$
is close to the energy gap $E_{sp}$, higher orbital bands are not
significantly populated in the driven process and the system is well
described by a simple two-band tight-binding model~\cite{Zheng2015}.
Following the standard procedure in Floquet theory, we obtain the effective
single-particle Hamiltonian
\begin{equation}
H_{0}^{\text{eff}}=\left(
\begin{array}{cc}
\epsilon _{s}(q_{x}) & \Delta _{sp} \\
\Delta _{sp}^{\ast } & \epsilon _{p}(q_{x})-|\Delta _{\omega }|%
\end{array}
\right) ,  \label{heff}
\end{equation}
where
\begin{equation}
\Delta _{sp}=-i[\Omega -\alpha \sin (q_{x}d)+\beta \cos (q_{x}d)]e^{-i\phi
_{0}}
\end{equation}
is the coupling between $s$- and $p$-orbital bands that is induced by the
moving lattice potential for $\Delta_\omega>0$~\cite{note1}, and $\epsilon
_{s}$ and $\epsilon _{p} $ are the energy dispersions for the uncoupled
orbital bands. The three coupling coefficients $\Omega$, $\alpha$ and $\beta$
are given by $\Omega =\frac{V^{\prime }}{4}\langle s_{i}|\sin
(2k_{L}x)|p_{i}\rangle $, $\alpha =\frac{V^{\prime }}{2}\langle s_{i}|\cos
(2k_{L}x)|p_{i+1}\rangle $ and $\beta =\frac{V^{\prime }}{2}\langle
s_{i}|\sin (2k_{L}x)|p_{i+1}\rangle $, where $|s_{i}\rangle $ and $%
|p_{i}\rangle $ are the maximally localized Wannier orbital states in the $i$%
-th site. $\Omega $ is the coupling between $s$- and $p$- orbital states in
the same lattice site, while $\alpha $ and $\beta $ are the couplings
between $s$- and $p$-orbital states of nearest neighbouring sites. SMC
between $s$-$p$ band psuedospins is represented by $\alpha \sin
(q_{x}d)\sigma _{x}$.

This derivation shows that the inversion symmetry of FB band structure is
broken due to the coexistence of couplings of different parities. When the
moving lattice depth is adiabatically ramped on, the quasimomentum of the
BEC gradually shifts away from $q_{x}=0$ in a definite direction following
the hybrid band minimum. This is quite different from previous shaken
lattice experiments~\cite{Parker2013} where the inversion symmetry of the
band was preserved and the BEC could spontaneously choose either side of $%
q_{x}=0$ as its ground state. In that case, the BEC needed to be accelerated
to break the inversion symmetry. In our scheme, the position of the true
minimum is uniquely determined by the moving velocity direction, moving
lattice depth, and driving frequency.

This minimal two-band model captures the essential physics of the driven
lattices as we have seen through the comparison of experimental measurements
and theoretical values (see Figs.~\ref{velocity} and \ref{strength}),
demonstrating the observation of SMC between $s$-$p$ band pseudospins.
However, this model may deviate from the experiment when the modulated
dynamics involve additional orbital bands or when the nonlinear interaction
is strong such that the single-particle band structure will be renormalized
by the interaction term.

\begin{figure}[t]
\centering\includegraphics[width=0.48\textwidth]{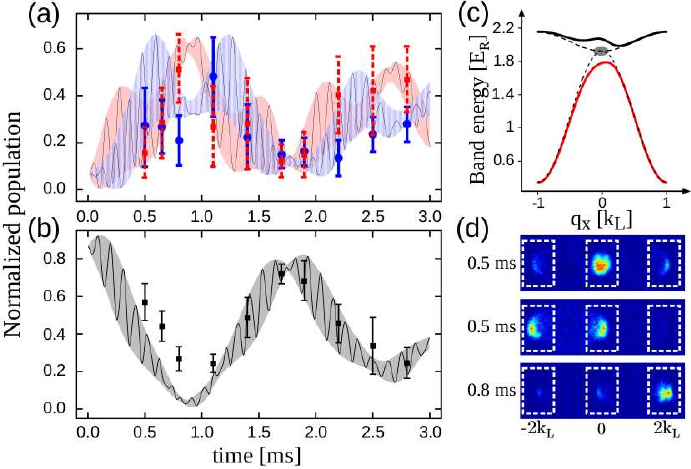}
\caption{\textbf{Quench dynamics after suddenly jumping on the coupling
between the $s$ and $p$ band.} (a, b) Normalized occupation of the momentum
component $-2 \hbar k$ (blue points with solid error bars) and $+2 \hbar k$
(red points with dashed error bars) in (a) and $0~\hbar k$ in (b). The dots
are the average of ten experimental measurements for each time. The error
bars indicate the spread of the experimental data. The shaded areas are the
results of numerical GP simulations calculated for a homogeneous
distribution of different initial phases $\protect\phi_0$. The black curve
represents the calculation for phase $\protect\phi_0=0$ (c) Bandstructure
plot. Jumping on the moving lattice places the BEC (black ellipse) into the
gap between two FB bands. (d) Experimental images taken $0.5~ms$ after the
quench for the top two images and $0.8~ms$ after the quench for the bottom
image. Dashed squares indicate the areas used for counting the atom number
in the $-2 \hbar k$ (left square), $0$ (middle square) and $+2 \hbar k$
component (right square). The sum of the atoms in all three boxes is used
for the normalization of the experimental data in panels (a) and (b).}
\label{phase}
\end{figure}

\noindent\textbf{Quench dynamics}. Since a Floquet system is generated by a
time-periodic Hamiltonian, an important question concerns the role of the
initial phase of the driving field~\cite{Goldman2014a}. For the system
considered in this work, this phase determines the relative positions
between the moving and static lattice sites. Though the relative phase does
not change the effective band structure (Eq.~\ref{heff}), and thus the
time-averaged dynamics, it can play a crucial role in the micromotion of the
BEC. To demonstrate the effect of the initial relative phase, we study the
oscillations in the population of the momentum components $k_{x}=0,\pm
2k_{L} $ after a quantum quench. Figure~\ref{phase} (a-b) present such
quench dynamics after adiabatically ramping on the static lattice to $%
5.47~E_{R}$ followed by a sudden jump on of the moving lattice to $%
V_{x}^{\prime }=1E_{R}$ with an on-resonant driving frequency~$|\Delta
_{\omega }|=E_{sp}$ (Fig.~\ref{phase} (c)). We focus on the evolution during
the first 3~ms, during which the BEC mainly stays at $q_{x}=0$ without
significant dipole motion in the hybrid bands. The symbols in Fig.~\ref%
{phase} (a-b) are experimental data averaged over ten measurements for each
time step. There is significant spread in the data for each time step, as
indicated by the vertical error bars. This spread is due to the initial
phase $\phi _{0}$ between the static and the moving lattice, which is
uncontrolled in the experiment, such that each iteration realizes a case
with a different, random $\phi _{0}$. The shaded areas represent the result
of numerical GP simulations for a homogeneous spread of relative phases. The
experimental error bars are in reasonable agreement with the expectation
based on these numerics. The numerics reveal that for a fixed initial phase
there are two oscillation periods of different timescales (Fig.~\ref{phase}%
(b)). The fast oscillation (of period $T\approx 0.1ms$) corresponds to the
micromotion of particles under the high-frequency periodic driving, whereas
the slow oscillation ($T\approx 1.75ms$) corresponds to the time-averaged
effective Rabi oscillations between the two hybrid FB bands. For longer
holding time, the periodicity is slightly broken due to a small dipole
motion.

\noindent \textbf{Discussion}. We have realized and characterized a new kind
of SMC with lattice bands as pseudospins. This not only provides a powerful
tool to control orbital states with a driving field, but also enriches the
study of novel quantum matter using hybrid orbital bands. There are many
directions that can be taken along this route, e.g., the engineering of
similar SMC in higher dimensional systems involving different orbital bands,
and quantitative analysis and measurements of the effects of strong
interactions on the effective bands. The realization of similar SMC for
fermionic atoms such as $^{6}$Li and $^{40}$K with tunable interactions may
open the door for exploring exotic quantum matters such as FFLO superfluids
and Majorana fermions.

\noindent \textbf{Acknowledgements} M.A.K, M.E.M., P.E. are supported by the
National Science Foundation (NSF) through Grant No. PHY-1306662. C. Qu and
C. Zhang are supported by ARO (W911NF-12-1-0334) and AFOSR
(FA9550-13-1-0045).

\noindent \textbf{Author contributions} M.A.K., C.Q., C.Z. and P.E.
conceived the experiment and theoretical modeling; M.A.K., M.E.M. and P.E.
performed the experiments; C.Q., C.Z. performed the theoretical
calculations; C.Z. and P.E. supervised the project.

\noindent \textbf{Competing financial interests}: The authors declare no
competing financial interests.\newline


\begin{thebibliography}{99}
\bibitem{note1} For $\Delta_\omega<0$, we have $\Delta _{sp}=i[\Omega
+\alpha \sin (q_{x}d)+\beta \cos (q_{x}d)]e^{i\phi _{0}}$. Note that $\beta$
is usually much smaller than $\Omega$ and $\alpha$, however we keep it for
completeness.
\end{thebibliography}
\end{document}